\documentstyle[12pt]{article}

\def\J{{\cal J}}  %%

\begin{document}

\title{Short-Range Spin Glasses: The Metastate Approach}

\author{C.~M. Newman\thanks{
Courant Institute of Mathematical Sciences, New York University, 
251 Mercer St., New York, NY 10012, USA.~
Research supported in part by
NSF Grant DMS-01-02587.} 
~~and ~D.~L. Stein\thanks{
Departments of Physics and Mathematics, University of Arizona, 
Tucson, AZ 85721, USA.~
Research supported in part by NSF Grant DMS-01-02541.}}

\date{}
\maketitle

\begin{abstract}
We discuss the {\it metastate\/}, a probability measure on thermodynamic
states, and its usefulness in addressing difficult questions pertaining to
the statistical mechanics of systems with quenched disorder, in particular
short-range spin glasses.  The possible low-temperature structures of
realistic (i.e., short-range) spin glass models are described, and a number
of fundamental open questions are presented.

\end{abstract}

\section{Introduction}
\label{sec:intro}

The nature of the low-temperature spin glass phase in short-range models
remains one of the central problems in the statistical mechanics of
disordered systems~\cite{BY86,Chowd86,MPV87,Stein89,FH91,Dotsenko01,NS03}.
While many of the basic questions remain unanswered, analytical and
rigorous work over the past decade have greatly streamlined the number of
possible scenarios for pure state structure and organization at low
temperature, and have clarified the thermodynamic behavior of these
systems.

The unifying concept behind this work is that of the {\it metastate\/}.  It
arose independently in two different constructions~\cite{AW90,NS96b}, which
were later shown to be equivalent~\cite{NSBerlin}.  The metastate is a
probability measure on the space of all thermodynamic states.  Its
usefulness arises in situations where multiple 
``competing'' pure states may be
present.  In such situations it may be difficult to construct
individual states in a measurable and canonical way; the
metastate avoids this difficulty by focusing instead on statistical
properties of the states.

An important aspect of the metastate approach is that it relates, by its
very construction~\cite{NS96b}, 
the observed behavior of a system in large but finite
volumes with its thermodynamic properties.  It therefore serves as a
(possibly indispensable) tool for analyzing and understanding both the
infinite-volume and finite-volume properties of a system, particularly in
cases where a straightforward interpolation between the two may be
incorrect, or their relation otherwise difficult to analyze.  

We will focus on the Edwards-Anderson~(EA) Ising spin glass
model~\cite{EA75}, although most of our discussion is relevant to a much
larger class of realistic models.  The EA model is described by the
Hamiltonian
\begin{equation} 
\label{eq:EA}
{\cal H}_{\cal J}= -\sum_{\langle x,y\rangle} J_{xy} \sigma_x \sigma_y\quad ,
\end{equation}
where ${\cal J}$ denotes a particular realization of all of the couplings
$J_{xy}$ and the brackets indicate that the sum is over nearest-neighbor
pairs only, with $x,y\in{\bf Z}^d$.  We will take Ising spins,
$\sigma_{x}=\pm 1$; although this will affect the details of our
discussion, it is unimportant for our main conclusions.  The couplings
$J_{xy}$ are quenched, independent, identically distributed random
variables whose common distribution $\mu$ is symmetric about zero.

\section{States and Metastates}
\label{sec:states}

We are interested in both finite-volume and infinite-volume Gibbs states.
For the cube of length scale $L$, $\Lambda_L= \{-L, -L+1, \cdots ,L\}^d$,
we define ${\cal H}_{{\cal J},L}$ to be the restriction of the EA
Hamiltonian to $\Lambda_L$ with a specified boundary condition (b.c.) such
as free or fixed or periodic.  Then the finite-volume Gibbs distribution
$\rho_{{\cal J}}^{(L)}= \rho_{{\cal J},\beta}^{(L)}$ on $\Lambda_L$ (at
inverse temperature $\beta = 1/T$) is:
\begin{equation}
\label{eq:finite}
%\rho_{{\cal J}^{(L)},\beta}
\rho_{{\cal J},\beta}^{(L)}(\sigma)=Z_{L}^{-1} \exp \{-\beta {\cal
H}_{{\cal J},L}(\sigma)\}\quad ,
\end{equation}
where the partition function $Z_{L}(\beta)$ is such that the sum of
$\rho_{{\cal J},\beta}^{(L)}$ over all $\sigma$ yields one.
(In this and all succeeding definitions, the dependence on spatial
dimension $d$ will be suppressed.)

{\it Thermodynamic\/} states are described by {\it infinite\/}-volume Gibbs
measures.  At fixed inverse temperature $\beta$ and coupling realization
${\cal J}$, a thermodynamic state~$\rho_{{\cal J},\beta}$ is the limit, as
$L\to\infty$, of some sequence of such finite-volume measures (each with a
specified b.c., which may remain the same or may change with $L$).  A
thermodynamic state $\rho_{{\cal J},\beta}$ can also be characterized
intrinsically through the Dobrushin-Lanford-Ruelle (DLR)
equations~\cite{Ge88}: for any $\Lambda_L$, the conditional distribution
of~$\rho_{{\cal J},\beta}$ (conditioned on the sigma-field generated by
$\{\sigma_x:x\in{\bf Z}^d\backslash\Lambda_L\}$ is $\rho_{{\cal
J},\beta}^{(L),{\overline\sigma}}$, where $\overline{\sigma}$ is given by
the conditioned values of $\sigma_x$ for $x$ on the boundary of
$\Lambda_L$.

Consider now the set ${\cal G}={\cal G}({\cal J},\beta)$ of all
thermodynamic states at a fixed $({\cal J},\beta)$.  The set of extremal,
or pure, Gibbs states is defined by
\begin{equation}
\label{eq:pure}
\mbox{ex }{\cal G} = {\cal G}\setminus \{a \rho_1+(1-a)\rho_2:~a
\in (0,1);~\rho_1,\rho_2\in {\cal G}; ~\rho_1\neq \rho_2 \}\, ,
\end{equation}
and the number of pure states ${\cal N}({\cal J},\beta)$ at $({\cal
J},\beta)$ is the cardinality $|\mbox{ex }{\cal G}|$ of $\mbox{ex }{\cal
G}$.  It is not hard to show that, in any $d$ and for a.e.~${\cal J}$, the
following two statements are true: 1) ${\cal N}=1$ at sufficiently low
$\beta>0$ ; 2) at any fixed $\beta$, ${\cal N}$ is constant a.s.~with
respect to the ${\cal J}$'s.  (The last follows from the measurability and
translation-invariance of ${\cal N}$, and the translation-ergodicity of
the disorder distribution of ${\cal J}$.)

A pure state $\rho_{\alpha}$ (where $\alpha$ is a pure state index) can
also be intrinsically characterized by a {\it clustering property\/}; for
two-point correlation functions this reads
\begin{equation}
\label{eq:clustering}
\langle\sigma_x\sigma_y\rangle_{\rho_{\alpha}} - \, 
\langle\sigma_x\rangle_{\rho_{\alpha}}\langle\sigma_y\rangle_{\rho_{\alpha}}
\to \, 0
\end{equation}
as $|x-y|\to\infty$.  A simple observation~\cite{NS92} with important
consequences for spin glasses, is that if many pure states exist, a
sequence of $\rho_{{\cal J},\beta}^{(L)}$'s, with b.c.'s and $L$'s chosen
independently of $\J$, will generally not have a (single) limit.  We call this
phenomenon {\it chaotic size dependence\/} (CSD).

We will be interested in the properties of $\mbox{ex }{\cal G}$ at low
temperature.  If the spin-flip symmetry present in the EA Hamiltonian
Eq.~(\ref{eq:EA}) is spontaneously broken above some dimension $d_0$ and
below some temperature $T_c(d)$, there will be at least a {\it pair\/} of
pure states, such that their even-spin correlations are identical, and
their odd-spin correlations have the opposite sign.  Assuming that such
broken spin-flip symmetry indeed exists for $d>d_0$ and $T<T_c(d)$, the
question of whether there exists {\it more than one\/} such pair (of
spin-flip related extremal infinite-volume Gibbs distributions) is a
central unresolved issue for the EA and related models.  If many such pairs
should exist, we can ask about the structure of their relations to one
another, and how this structure would manifest itself in large but finite
volumes.

To do this we use an approach, introduced in Ref.~{\cite{NS96b}, to
studying inhomogeneous and other systems with many competing pure states.
This approach, based on an analogy to chaotic dynamical systems, requires
the construction of a new thermodynamic quantity which we call the {\it
metastate,\/} which is a probability measure $\kappa_{\cal J}$ on the
thermodynamic states.  The metastate allows an understanding of CSD by
analyzing the way in which $\rho_{{\cal J},\beta}^{(L)}$ ``samples'' from
its various possible limits as $L\to\infty$.

The analogy with chaotic dynamical systems can be understood as follows.
In dynamical systems, the chaotic motion along a deterministic orbit is
analyzed in terms of some appropriately selected probability measure,
invariant under the dynamics. Time along the orbit is replaced, in our
context, by $L$ and the phase space of the dynamical system is replaced by
the space of Gibbs states.

In~\cite{NS96b}, we considered (as always, at fixed $\beta$, which will
hereafter be suppressed for ease of notation) a `microcanonical ensemble'
$\kappa_N$ in which each of the finite-volume Gibbs states $\rho_{{\cal
J}}^{(L_1)},\rho_{{\cal J}}^{(L_2)},\ldots, \rho_{{\cal J}}^{(L_N)}$ has
weight $N^{-1}$.  The ensemble $\kappa_N$ converges to a metastate 
$\kappa_{\cal J}$ as $N\to\infty$, in the following sense: for every (nice)
function $g$ on states (e.g., a function of finitely many correlations),
\begin{equation}
\label{eq:macro}
\lim_{N\to\infty}N^{-1}\sum_{{\ell}=1}^{N} g(\rho^{({L_{\ell}})}) =
\int g(\Gamma)\, d \kappa_{\cal J}(\Gamma) \quad .
\end{equation}
The information contained in $\kappa_{\cal J}$ 
effectively specifies the fraction of
cube sizes $L_{\ell}$ which the system spends in different (possibly mixed)
thermodynamic states $\Gamma$ as $\ell \to \infty$.

A different, but in the end equivalent approach, 
based on ${\cal J}$-randomness, is due to Aizenman
and Wehr~\cite{AW90}.  Here one considers the random pair $({\cal
J},\rho_{\cal J}^{(L)})$, defined on the underlying probability space of
${\cal J}$, and takes the limit $\kappa^\dagger$ (with conditional
distribution $\kappa^{\dagger}_{\cal J}$, given ${\cal J}$), via finite
dimensional distributions along some subsequence.  We omit details here,
and refer the reader to~\cite{AW90, NSBerlin}.  We note, though, the
important result that a {\it deterministic\/} subsequence of volumes can be
found on which both (\ref{eq:macro}) is valid 
and $({\cal J},\rho_{\cal J}^{(L)})$
converges,  with $\kappa^{\dagger}_{\cal J} = \kappa_{\cal
J}$~\cite{NSBerlin}.

In what follows we use the term ``metastate'' as shorthand for the
$\kappa_{\cal J}$ constructed using periodic b.c.'s on a sequence of
volumes chosen independently of the couplings, and along which
$\kappa_{\cal J}=\kappa^\dagger_{\cal J}$.  We choose periodic b.c.'s for
specificity; the results and claims discussed are expected to be
independent of the boundary conditions used, as long as they are chosen
independently of the couplings.

\section{Low Temperature Structure of the EA Model}
\label{sec:lowtemp}

There have been several scenarios proposed for the spin glass phase of the
Edwards-Anderson model at sufficiently low temperature and high dimension.
These remain speculative, because it has not even been proved that a phase
transition from the high-temperature phase exists at positive temperature
in {\it any\/} finite dimension.  

We noted earlier that, at sufficiently high temperature in any dimension
(and at all nonzero temperatures in one and presumably two dimensions,
though the latter assertion has not been proved), there is a unique Gibbs
state.  It is conceivable that this remains the case in all dimensions and
at all nonzero temperatures, in which case the metastate $\kappa_{\cal J}$
is, for a.e.~${\cal J}$, supported on a single, pure Gibbs state
$\rho_{\cal J}$.  (It is important to note, however, that in principle such
a trivial metastate could occur even if ${\cal N}> 1$; indeed, just such a
situation of ``weak uniqueness''~\cite{EnF85,COE87} happens in 
very long range
spin glasses at high temperature~\cite{FZ87,GNS93}.)

A phase transition has been proved to exist~\cite{ALR87} in the
Sherrington-Kirkpatrick~(SK) model~\cite{SK75}, which is the
infinite-range version of the EA model.
Numerical~\cite{BY86,O85,OM85,KY96} and some analytical~\cite{FS90,TH96}
work has led to a general consensus that above some dimension (typically
around three or four) there does exist a positive temperature phase
transition below which spin-flip symmetry is broken, i.e., in which pure
states come in pairs, as discussed below Eq.~(\ref{eq:clustering}).
Because much of the literature has focused on this possibility, we assume
it in what follows, and will see that the metastate approach is highly
useful in restricting the scenarios that can occur.

The simplest such scenario is a two-state picture in which, below the
transition temperature $T_c$, there exists a single pair of global
flip-related pure states $\rho_{\cal J}^\alpha$ and $\rho_{\cal
J}^{-\alpha}$.  In this case there is no CSD for periodic
b.c.'s and the metastate can be written
\begin{equation}
\label{eq:dsmeta}
\kappa_{\cal J}=\delta_{{1\over 2}\rho_{\cal J}^\alpha+{1\over 2}\rho_{\cal
J}^{-\alpha}}\, .
\end{equation}
That is, the metastate is supported on a single (mixed) thermodynamic
state.

The two-state scenario that has received the most attention in the
literature is the {\it droplet/scaling} picture~\cite{Mac,BM,FH}.  In this
picture a low-energy excitation above the ground state in $\Lambda_L$ is a
droplet whose surface area scales as $l^{d_s}$, with $l\sim O(L)$ and
$d_s<d$, and whose surface energy scales as $l^\theta$, with $\theta>0$ (in
dimensions where $T_c>0$).  More recently, an alternative picture has
arisen~\cite{KM00,PY00} in which the low-energy excitations differ from
those of droplet/scaling in that their energies scale as $l^{\theta'}$, with
$\theta'=0$.

The low-temperature picture that has perhaps generated the most attention
in the literature is the replica symmetry breaking~(RSB)
scenario~\cite{BY86,Dotsenko01,MPR94,MPR97,FMPP98,MPRRZ00,MP00a,MP00b},
which assumes a rather complicated pure state structure, inspired by
Parisi's solution of the SK~model~\cite{MPV87,Pa79,Pa83,MPSTV84}.  This is
a many-state picture (${\cal N}=\infty$ for a.e.~${\cal J}$) in which the
ordering is described in terms of the {\it overlaps\/} between states.
There has been some ambiguity in how to describe such a picture for
short-range models; we start with the prevailing, or standard, view.
Consider any reasonably constructed thermodynamic state $\rho_{\cal J}$
(see~\cite{NSBerlin} for more details) --- e.g, the {\it average\/} over
the metastate $\kappa_{\cal J}$
\begin{equation}
\label{eq:metaav}
\rho_{\cal J}=\int \Gamma \, d\kappa_{\cal J}(\Gamma) \quad .
\end{equation}

Now choose $\sigma$ and $\sigma'$ from the product distribution $\rho_{\cal
J}(\sigma)\rho_{\cal J}(\sigma')$.  The overlap $Q$ is defined as
\begin{equation}
\label{eq:overlap}
Q=\lim_{L\to\infty}|\Lambda_L|^{-1}\sum_{x\in\Lambda_L}\sigma_x\sigma'_{x}\, ,
\end{equation}
and $P_{\cal J}(q)$ is defined to be its probability distribution.

In the standard RSB picture $\rho_{\cal J}$ is a mixture of infinitely many
pure states, each with a specific ${\cal J}$-dependent weight $W$:
\begin{equation}
\label{eq:sum}
\rho_{\cal J}(\sigma)=\sum_\alpha W_{\cal J}^\alpha\rho_{\cal J}^\alpha (\sigma)\ .
\end{equation}
If $\sigma$ is drawn from $\rho_{\cal J}^\alpha$ and $\sigma'$
from $\rho_{\cal J}^\beta$, then the expression in Eq.~(\ref{eq:overlap}) equals 
its thermal mean,
\begin{equation}
\label{eq:qab}
q_{\cal J}^{\alpha\beta}=\lim_{L\to\infty}|\Lambda_L|^{-1}
\sum_{x\in\Lambda_L} \langle\sigma_x\rangle_\alpha
\langle\sigma_x\rangle_\beta \quad ,
\end{equation}
and hence $P_{\cal J}$ is given by
\begin{equation}
\label{eq:PJ(q)}
P_{\cal J}(q)=\sum_{\alpha,\beta}W_{\cal J}^\alpha W_{\cal J}^\beta
\delta(q-q_{\cal J}^{\alpha\beta})\quad .
\end{equation}
The {\it self-overlap\/}, or EA order parameter, is given by
$q_{EA}=q_{\cal J}^{\alpha\alpha}$ and (at fixed 
$T$) is thought to be
independent of both $\alpha$ and ${\cal J}$ (with probability one).

According to the standard RSB scenario, the $W_{\cal J}^\alpha$'s and
$q_{\cal J}^{\alpha\beta}$'s are non-self-averaging (i.e., ${\cal
J}$-dependent) quantities, except for $\alpha=\beta$ or its global flip,
where $q_{\cal J}^{\alpha\beta}=\pm q_{EA}$.  The average $P_s(q)$ of
$P_{\cal J}(q)$ over the disorder distribution of ${\cal J}$ is
predicted to be a mixture of two delta-function components at $\pm q_{EA}$
and a continuous part between them.

However, it was proved in~\cite{NS96a} that this scenario cannot occur,
because of the translation-invariance of $P_{\cal J}(q)$ and the
translation-ergodicity of the disorder
distribution.  Neverthelesss, the metastate approach
suggests an alternative, nonstandard, RSB scenario, which we now describe.

The idea behind the nonstandard RSB picture (referred to by us as the
nonstandard SK picture in earlier papers) is to produce the finite-volume
behavior of the SK model to the maximum extent possible.  We
therefore assume in this picture that in each $\Lambda_{L}$, the
finite-volume Gibbs state $\rho_{\cal J}^{(L)}$ is well approximated deep
in the interior by a mixed thermodynamic state $\Gamma^{({L})}$,
decomposable into many pure states $\rho_{\alpha_{L}}$ (explicit dependence
on ${\cal J}$ is suppressed for ease of notation). More precisely, each
$\Gamma$ in $\kappa_{\cal J}$ satisfies
\begin{equation} 
\label{eq:gamma}
\Gamma\, =\sum_{\alpha_{\Gamma}}W_{\Gamma}^{\alpha_{\Gamma}}\rho_{\alpha_{\Gamma}}\,
\end{equation} 
and is presumed to have a nontrivial overlap distribution for
$\sigma,\sigma'$ from $\Gamma(\sigma)\Gamma(\sigma')$:
\begin{equation} 
\label{eq:gamov}
P_{\Gamma}(q)=\sum_{\alpha_{\Gamma},\beta_{\Gamma}}W_{\Gamma}^{\alpha_{\Gamma}}
W_{\Gamma}^{\beta_{\Gamma}}\delta(q-q_{\alpha_{\Gamma}\beta_{\Gamma}})
\end{equation} 
as did $\rho_{\cal J}$ in the standard RSB picture.

Because $\kappa_{\cal J}$, like its counterpart $\rho_{\cal J}$ in the
standard SK picture, is translation-covariant, the resulting {\it
ensemble\/} of overlap distributions $P_{\Gamma}$ is independent of ${\cal
J}$.  Because of the CSD present in this scenario, the overlap distribution
for $\rho_{\cal J}^{(L)}$ varies with $L$, no matter how large ${L}$
becomes.  So, instead of averaging the overlap distribution over ${\cal
J}$, the averaging must now be done over the states $\Gamma$ {\it within
the metastate\/} $\kappa_{\cal J}$, all at fixed ${\cal J}$:
\begin{equation}
\label{eq:Pgamma} 
P_{ns}(q)= \int P_\Gamma(q)\kappa_{\cal J}(\Gamma)d\Gamma\, .
\end{equation} 
The $P_{ns}(q)$ is the same for a.e.~${\cal J}$, and has a form analogous
to the $P_s(q)$ in the standard RSB picture.

However, the nonstandard RSB scenario seems rather unlikely to occur in any
natural setting, because of the following result:

{\bf Theorem~\cite{NS98}.}  Consider two metastates constructed along (the
same) deterministic sequence of $\Lambda_L$'s, using two different
sequences of flip-related, coupling-independent b.c.'s (such as periodic
and antiperiodic).  Then with probability one, these two metastates are the
same.

The proof is given in~\cite{NS98}, but the essential idea can be easily
described here.  As discussed in Sec.~\ref{sec:states}, $\kappa_{\cal
J}=\kappa^\dagger_{\cal J}$; but $\kappa^\dagger_{\cal J}$ is constructed
by a limit of finite dimensional distributions, which means averaging over
other couplings including ones near the system boundary, and hence gives
the same metastate for two flip-related boundary conditions.

This invariance with respect to different sequences of periodic and
antiperiodic b.c.'s means essentially that the frequency of appearance of
various thermodynamic states $\Gamma^{(L)}$ in finite volumes $\Lambda_L$
is {\it independent\/} of the choice of boundary conditions.  Moreover,
this same invariance property holds among any two sequences of {\it
fixed\/} boundary conditions (and the fixed boundary condition of choice
may even be allowed to vary arbitrarily along any single sequence of
volumes)!  It follows that, with respect to changes of boundary conditions,
the metastate is extraordinarily robust.

This should rule out all but the simplest overlap structures, and in
particular the nonstandard RSB and related pictures (for a full discussion,
see~\cite{NS98}).  It is therefore natural to ask whether the property of
metastate invariance allows {\it any\/} many-state picture.

There is one such picture, which we have called {\it chaotic pairs\/}, and
which is fully consistent with metastate invariance (our belief is that it
is the {\it only\/} many-state picture that fits naturally and easily into
results obtained about the metastate.)

Here the periodic b.c.~metastate is supported on infinitely many pairs of
pure states, but instead of Eq.~(\ref{eq:gamma}) one has
\begin{equation}
\label{eq:cp}
\Gamma \, =(1/2)\rho_{\alpha_{\Gamma}} + (1/2) \rho_{-\alpha_{\Gamma}}\, ,
\end{equation}
with overlap
\begin{equation}
\label{eq:cpov}
P_{\Gamma}=(1/2)\delta(q-q_{EA})+(1/2)\delta(q+q_{EA})\, .
\end{equation}

So there is CSD in the states but not in the overlaps, which have the same
form as a two-state picture in every volume.  The difference is that, while
in the latter case, one has the {\it same\/} pair of states in every
volume, in chaotic pairs the pure state pair varies chaotically as volume
changes.  If chaotic pairs is to be consistent with metastate invariance in
a natural way, then the number of pure state pairs should be {\it
uncountable\/}.  This allows for a `uniform' distribution (within the
metastate) over all of the pure states, and invariance of the metastate
with respect to boundary conditions could follow naturally.

\section{Open Questions}
\label{sec:open}

We have discussed how the metastate approach to the EA spin glass has
narrowed considerably the set of possible scenarios for low-temperature
ordering in any finite dimension, should broken spin-flip 
symmetry occur.
The remaining possibilities are either a two-state scenario, like
droplet/scaling, or the chaotic pairs picture if there exist many pure
states at some $(\beta,d)$.  Both have simple overlap structures.  The
metastate approach appears to rule out more complicated scenarios like RSB,
in which the approximate pure state decomposition in a typical large,
finite volume is a nontrivial mixture of many pure state pairs.

Of course, this doesn't answer the question of which, if either, of the
remaining pictures actually does occur in real spin glasses.  In this
section we list a number of open questions relevant to the above discussion.

{\bf Open Question 1.} Determine whether a phase transition occurs in any
finite dimension greater than one.  If it does, find the lower critical
dimension.

Existence of a phase transition does not necessarily imply two or more pure
states below $T_c$.  It could happen, for example, that in some dimension
there exists a single pure state at all nonzero temperatures, with
two-point spin correlations decaying exponentially above $T_c$ and more
slowly (e.g., as a power law) below $T_c$.  This leads to:

{\bf Open Question 2.} If there does exist a phase transition above some
lower critical dimension, determine whether the low-temperature spin glass
phase exhibits broken spin-flip symmetry.

If broken symmetry does occur in some dimension, then of course an obvious
open question is to determine the number of pure state pairs, and hence the
nature of ordering at low temperature.  A (possibly) easier question (but
still very difficult), and one which does not rely on knowing whether a
phase transition occurs, is to determine the zero-temperature --- i.e.,
ground state --- properties of spin glasses as a function of
dimensionality.  A ground state is an infinite-volume spin configuration
whose energy (governed by Eq.~(\ref{eq:EA})) cannot be lowered by flipping
any finite subset of spins.  That is, all ground state spin configurations
must satisfy the constraint
\begin{equation}
\label{eq:loop}
\sum_{\langle x,y\rangle\in {\cal C}}J_{xy}\sigma_x \sigma_y \ge 0
\end{equation}
along any closed loop ${\cal C}$ in the dual lattice.  

{\bf Open Question 3.}  How many ground state pairs is the $T=0$ periodic
boundary condition metastate supported on, as a function of $d$?

The answer is known to be one for $1D$, and a partial
result~\cite{NS2D00,NSCMP01} points towards the answer being one for $2D$
as well.  There are no rigorous, or even heuristic (except based on
underlying {\it ans\"atze\/}) arguments in higher dimension.

An interesting --- but unrealistic --- spin glass model in which the ground
state structure can be exactly solved (although not yet completely
rigorously) was proposed by the authors~\cite{NS94,NS96c} (see
also~\cite{BCM94}).  This ``highly disordered'' spin glass is one in which
the coupling magnitudes scale nonlinearly with the volume (and so are no
longer distributed independently of the volume, although they remain
independent and identically distributed for each volume).  The model
displays a {\it transition\/} in ground state multiplicity: below eight
dimensions, it has only a single pair of ground states, while above eight
it has uncountably many such pairs.  The mechanism behind the transition
arises from a mapping to invasion percolation and minimal spanning
trees~\cite{LB80,CKLW82,WW83}: the number of ground state pairs can be
shown to equal $2^{\cal N}$, where ${\cal N} = {\cal N}(d)$ is the number
of distinct global components in the``minimal spanning forest.''  The
zero-temperature free boundary condition metastate above eight dimensions
is supported on a uniform distribution (in a natural sense) on uncountably
many ground state pairs.

Interestingly, the high-dimensional ground state multiplicity in this model
can be shown to be {\it unaffected\/} by the presence of frustration,
although frustration still plays an interesting role: it leads to the
appearance of chaotic size dependence when free 
conditions are used.

Returning to the more difficult problem of ground state multiplicity in the
EA model, we note as a final remark that there could, in principle, exist
ground state pairs that are not in the support of metastates generated
through the use of coupling-independent boundary conditions.  If such
states exist, they may be of some interest mathematically, but are not
expected to play any significant physical role.  A discussion of these
putative ``invisible states'' is given in \cite{NS03}.

{\bf Open Question 4.}  If there exists broken spin-flip symmetry at a
range of positive temperatures in some dimensions, then what is the number
of pure state pairs as a function of $(\beta,d)$?

Again, the answer to this is not known above one dimension; indeed, the
prerequisite existence of spontaneously broken spin flip symmetry has not
been proved in any dimension.  A speculative paper by the
authors~\cite{NS01}, using a variant of the highly disordered model,
suggests that there is at most one pair of pure states in the EA model
below eight dimensions; but no rigorous arguments are known at this
time.

\end{document}